\documentclass[]{spie}  
\usepackage{wrapfig}
\usepackage{xcolor}
\usepackage{ulem}

\usepackage{amsmath,amsfonts,amssymb}
\usepackage{graphicx}
\usepackage[colorlinks=true, allcolors=blue]{hyperref}

\title{Precise Blaze Angle Measurements of Lithographically Fabricated Silicon Immersion Gratings}

\author[a]{Emily Lubar}
\author[a]{Daniel T. Jaffe}
\author[a]{Cynthia Brooks}
\author[b]{Sierra Hickman}
\author[a]{Michael Gully-Santiago}
\author[a]{Gregory Mace}

\affil[a]{University of Texas at Austin}
\affil[b]{University of Canterbury}

\begin{document} 
\maketitle

\begin{abstract}
Silicon immersion gratings and grisms enable compact, near-infrared spectrographs with high throughput. These instruments find use in ground-based efforts to characterize stellar and exoplanet atmospheres, and in space-based observatories. Our grating fabrication technique uses x-ray crystallography to orient silicon parts prior to cutting, followed by lithography and wet chemical etching to produce the blaze. This process takes advantage of the crystal structure and relative difference in etching rates between the (100) and (111) planes such that we can produce parts that have surface errors $<\frac{\lambda}{4}$. Previous measurements indicate that chemical etching can yield a final etched blaze that slightly differs from the orientation of the (111) plane. This difference can be corrected by the mechanical mount in the case of the immersion gratings, but doing so may compromise grating throughput due to shadowing. In the case of the grisms, failure to take the actual blaze into account will reduce the brightness of the undeviated ray. We report on multiple techniques to precisely measure the blaze of our in-house fabricated immersion gratings. The first method uses a scanning electron microscope to image the blaze profile, which yields a measurement precision of 0.5 degrees. The second method is an optical method of measuring the angle between blaze faces using a rotation stage, which yields a measurement precision of 0.2 degrees. Finally, we describe a theoretical blaze function modeling method, which we expect to yield a measurement precision of 0.1 degrees. With these methods, we can quantify the accuracy with which the wet etching produces the required blaze and further optimize grating and grism efficiencies.

\end{abstract}


\section{INTRODUCTION}
\label{sec:intro}  

Immersion gratings and grisms take advantage of the high index of refraction of silicon for infrared wavelengths, which increases the dispersion angle of reflected light relative to that of a free-space reflection grating by a factor of $n_{NIR}\sim3.4$. For a desired resolving power, immersion gratings allow for broad spectral grasp in a compact and inexpensive instrument design. The Immersion GRating Infrared Spectrometer (IGRINS) uses an immersion grating as the primary disperser\cite{10.1117/12.317263,10.1117/12.857164} to simultaneously cover H and K band at high resolution (R $\sim$ 45,000). Immersion gratings and grisms have found application in both ground-based (FIRST\cite{10.1117/12.2057023}, IGRINS\cite{10.1117/12.317263,10.1117/12.857164}, iSHELL\cite{10.1117/12.2232064}, MagNIFIES/GMTNIRS\cite{10.1117/12.2232994}) and space-based (FORCAST on SOFIA\cite{10.1117/12.857127}, NIRcam on JWST\cite{10.1117/12.857568}, TROPOMI-SWIR \cite{10.1117/12.869018, 10.1117/12.2309092}) instrumentation. The wavelength coverage and high-resolution combination allow us to access higher-order information about early stellar evolution and exoplanetary atmospheric characterization. \cite{Sokal_2020, Flagg_2019, Johns_Krull_2016} 

\begin{figure}
\centering
\includegraphics[width=0.8\textwidth]{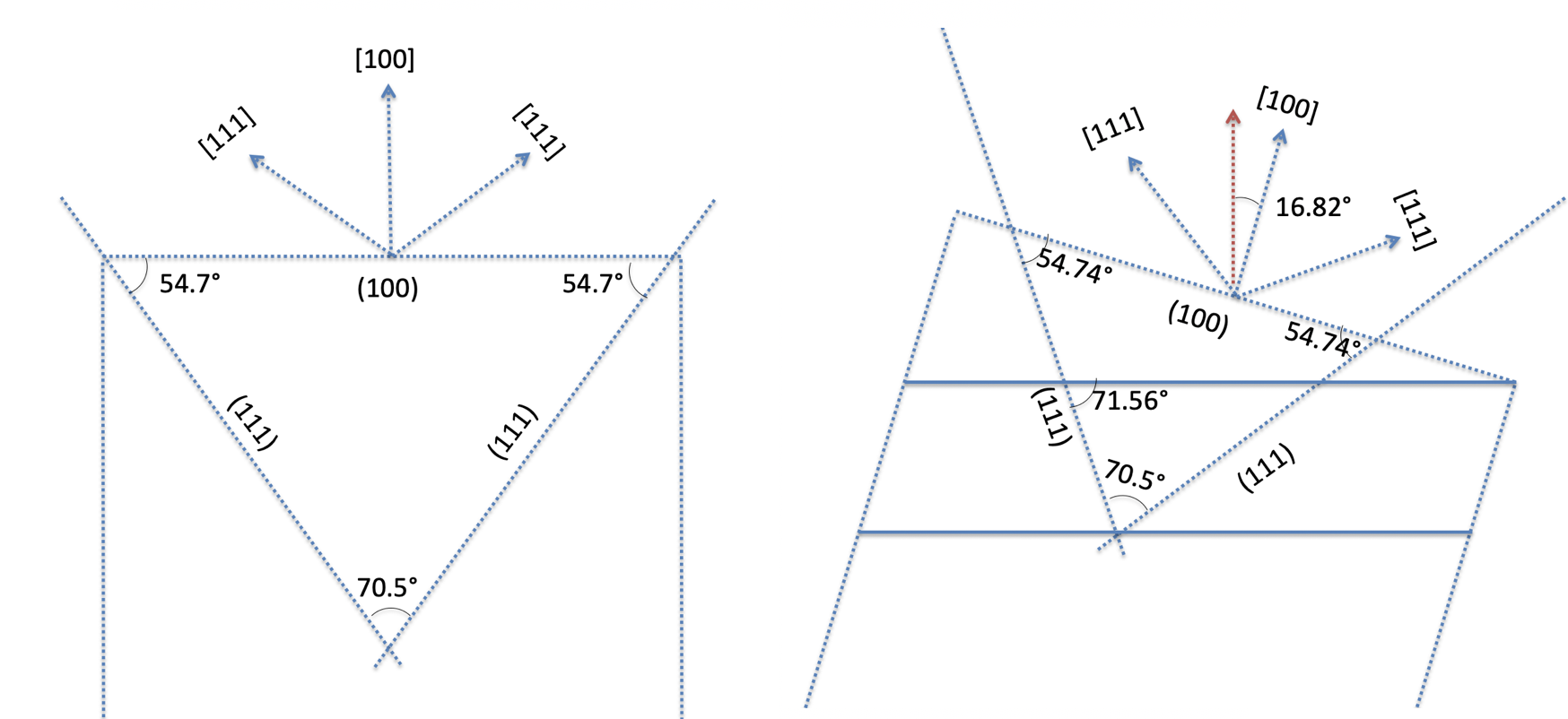}
\caption{\label{fig:fabricationAnd} The left diagram shows the alignment of the (100) and (111) planes in the silicon. On the right shows the same geometry, but tilted to the angle we cut the part from the silicon boule. The desired blaze is $\sim$71.56 degrees (also known as the industry standard R3) which we achieve by cutting the original slab of silicon along the thicker solid lines shown above. The final blaze angle is encoded by the orientation of the (111) crystalline plane and the desired cut.}
\end{figure}

Our in-house fabrication of immersion gratings uses a contact lithography and potassium hydroxide (KOH) wet-etch technique that takes advantage of the crystal structure of silicon. The grating blaze is defined when the part is initially cut from the larger silicon boule. As shown in Fig \ref{fig:fabricationAnd}, a cut is made relative to the (111) planes in the crystalline structure to define the target blaze angle. The part is aligned with x-ray crystallography (accurate to $\sim$0.05 degrees) to orient the boule prior to cutting. After the part is cut away from the boule, a layer of photoresist is deposited on the part surface. Using a quartz/chrome mask and collimated UV illumination system, the rows of the mask are imprinted onto the part to define the grating grooves\cite{10.1117/12.857164}. The grating is then put through a chemical bath which etches the grooves as defined by the imprinted mask pattern. These techniques produce parts that have surface errors $<\frac{\lambda}{4}$. For further detail on the fabrication process, see Marsh et al. 2006\cite{10.1117/12.672198}, Brooks et al. 2014 \cite{10.1117/12.2057468}, and Kidder et al. 2018 \cite{10.1117/12.2314271}.

The final blaze angle is defined by both the crystalline structure and the relative etch rate between the (100) and (111) planes. Previous measurements indicate that chemical etching can yield a final etched blaze that differs from the expected value by approximately 0.5 to 1 degrees. If the blaze of the grating is off, this can result in power deviating from the desired order and therefore a loss of throughput. For example, for a grating not used in immersion with an intended blazed of $\theta = 71.56$ degrees, but a true blaze of $\theta = 71.60$ degrees, there would be a $\sim 1\%$ loss in intensity in the blazed order (Fig \ref{fig:offset} illustrates this).

In addition, any error in the true blaze is magnified when the optic is used in immersion because of Snell's law; for the light to take a straight, uncorrected path across the interface of two different indices of refraction, it must be normal to the interface surface. This configuration is achieved in the instrument when the exit face of the grating is cut exactly parallel to the blaze of the grating. If the grating blaze is off, these surfaces will no longer be parallel and the light will no longer be normal to the cut surface after interacting with the grating surface. This discrepancy between the true blaze and the expected blaze (which defines the angle at which the entrance face is cut) deviates the light away from the expected path by a factor of 3.4, magnifying the problem. In some cases, such a discrepancy can be partially corrected by modifications to the optic's mount, but this can compromise efficiency when the modification introduces groove shadowing. Precisely characterizing the blaze produced by the wet-etch process allows us to make the final cut parallel to the true blaze, alleviating the need to make corrections after the fact. In the following sections, we present three methods for precisely determining the final blaze of our immersion gratings.

\section{Three Methods of blaze determination}

The following methods can be useful in any situation where the blaze angle is imprecisely known, or if one wishes to verify specifications provided by a vendor if the part is off the shelf or made custom. Each method outlined here yields a different level of precision.

\begin{figure}[t]
\centering
\includegraphics[width=0.6\textwidth]{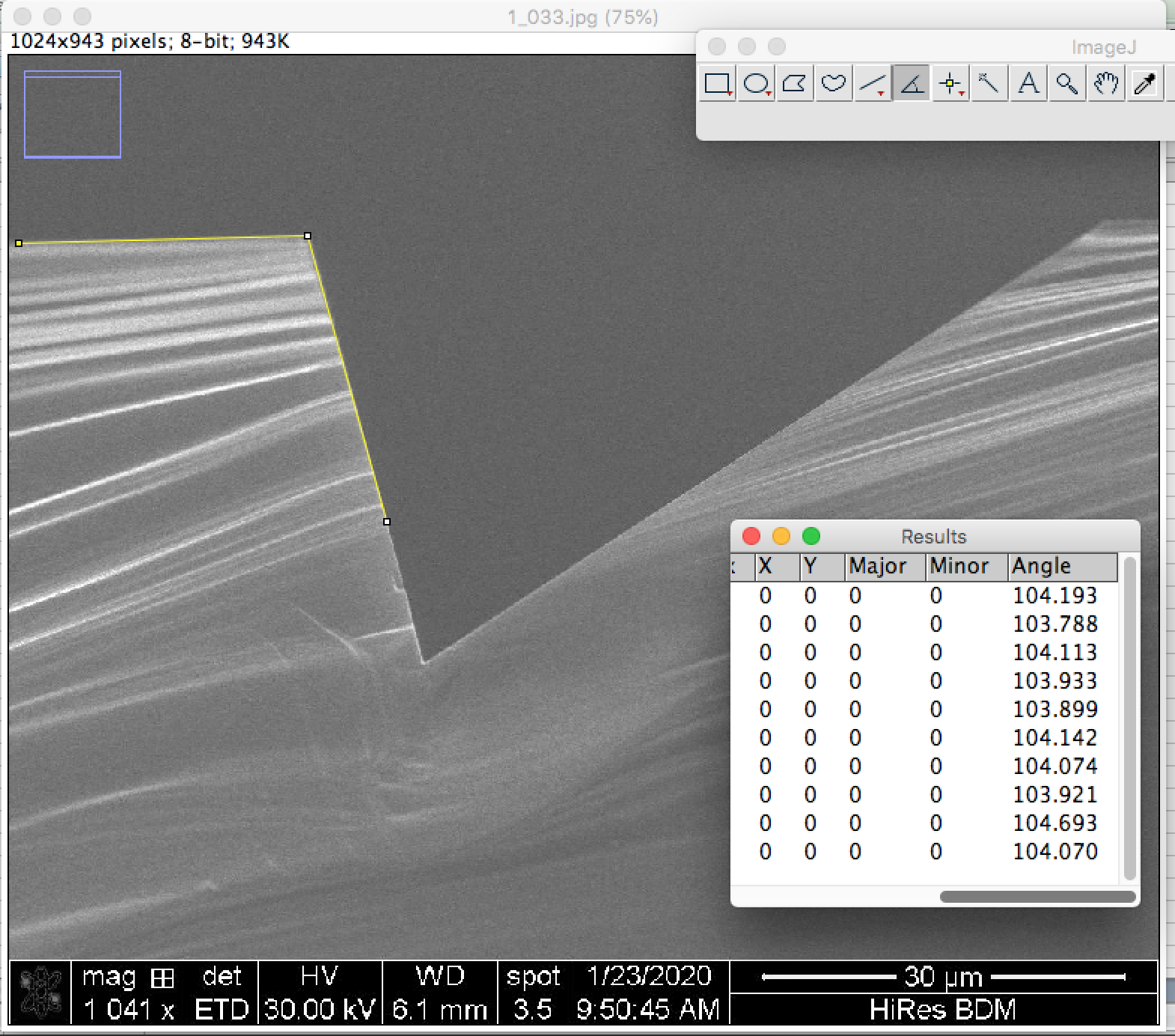}
\caption{\label{fig:SEMimg} The grating profile is imaged with an SEM. Using ImageJ software, the supplementary angle is measured multiple times and averaged (shown above by the yellow traced edges), then subtracted from 180 to obtain the final blaze measurement. This particular part was fabricated with a target blaze of R4. The precision of the measurement shown here is $\Delta x = \frac{R}{2}= \frac{x_{max}-x_{min}}{2} = 0.45$.}
\end{figure} 

\subsection{Method 1}
\label{sec:title}

The first method is to image a cross-section of the grating profile with a Scanning Electron Microscope (SEM), which yields a measurement precision of $\sim$0.5 degrees. After obtaining quality images with the SEM, measurements are made with the ImageJ\cite{IMGJ} software as shown in Fig.~\ref{fig:SEMimg}. ImageJ has a measurement tool that allows the user to trace lines on the image to obtain physical measurements. This method relies on the focus of the image and on the user correctly tracing out the edges of the grating profile that define the blaze angle. The primary limiting factor to the precision of this measurement method is the challenge associated with manually selecting line edges in a repeatable way. We trace out the profile to measure the supplement angle 10-15 times and take the average as our final value. The precision of this small N data set can be accurately described by dividing the range ($x_{max}-x_{min}$) by 2.

This blaze-measurement method is not always convenient for our purposes. In the intermediate stages of the fabrication process, the grating pattern is an island on the silicon part surface such that there are no exposed cross-sections on the part edges. Only after the part has been cut to its final deployment shape do we have an exposed cross-section to image. By the time this cut has been made, a part must already have been vetted and chosen for deployment in an instrument. Ideally, we want to measure the blaze precisely before resources are spent on this final cut, so it is necessary to have other methods of measuring the blaze that happens earlier in the fabrication process. For other applications, improving this measurement method might include making an automated program that finds and draws the lines that define the blaze angle. This improvement would then only be limited by the degree of focus of the SEM imaged profile (any section of the grating which does not lie in the plane of focus introduces some uncertainty; this effect can be seen along the top of the grating profile in Fig \ref{fig:SEMimg}). Rather than working to improve this method, we utilize higher precision methods which can be utilized earlier in the fabrication process. Methods 2 and 3 serve this purpose. 

\begin{figure}
\centering
\includegraphics[width=0.7\textwidth]{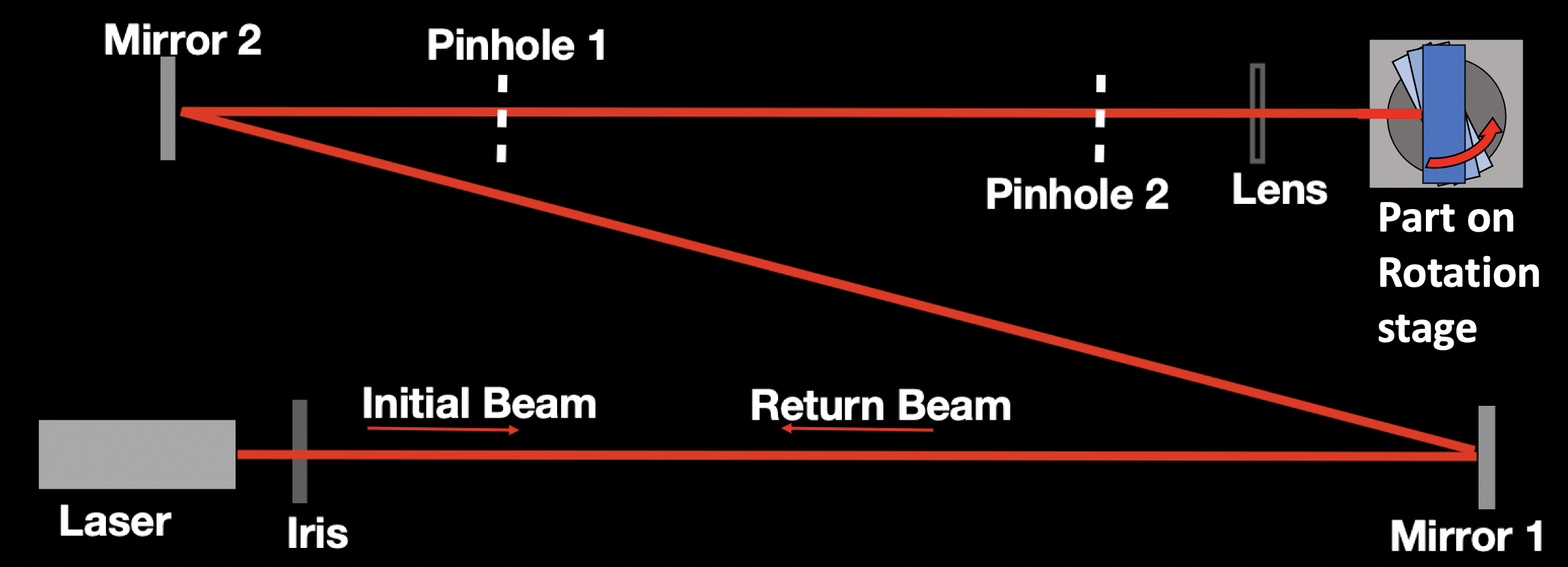}
\caption{\label{fig:setup}A top view of the optical set up used to probe the blaze angle in measurement method \#2. Pinholes are used for alignment, then removed for measurement process. The part is shown on top of the rotation stage, which allows for the rotated configurations shown in Fig \ref{fig:measurementwalls}.}
\end{figure} 

\begin{figure}
\centering
\includegraphics[width=0.7\textwidth]{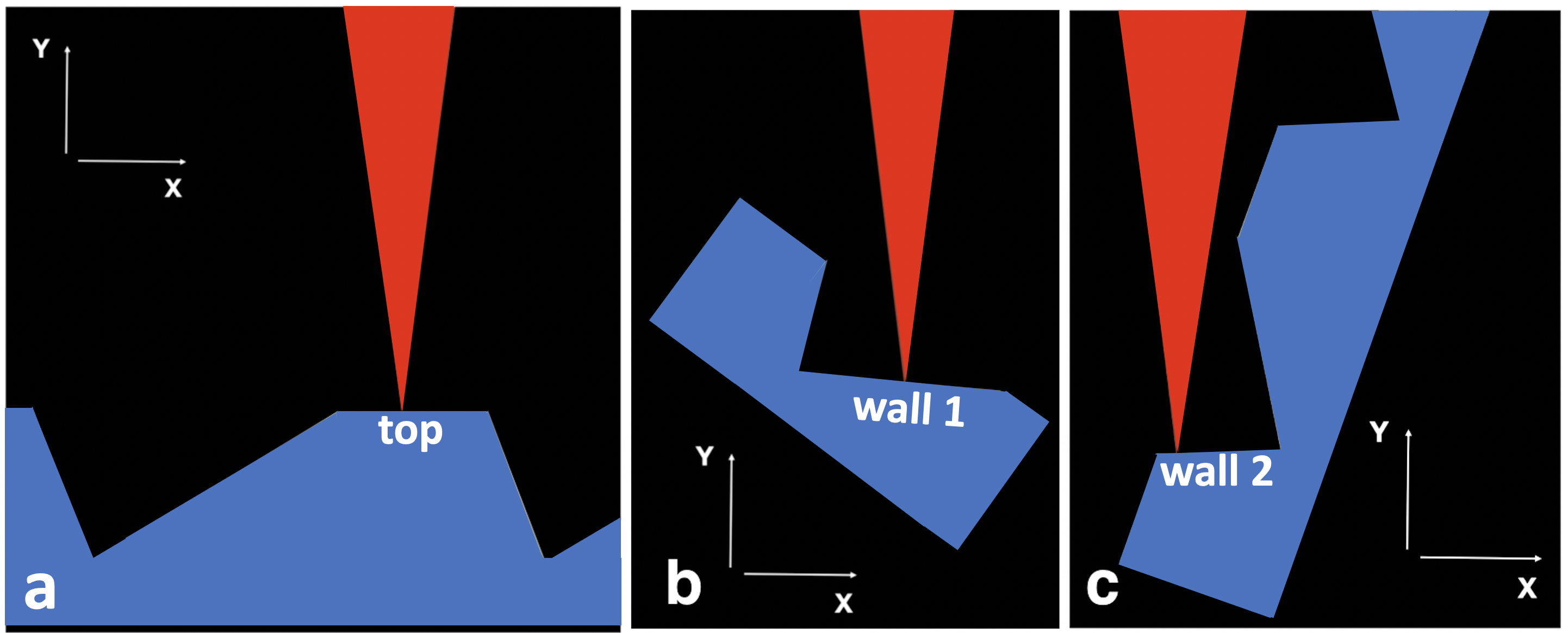}
\caption{\label{fig:measurementwalls} A top view of the grating surface as it is rotated on the rotation stage. Figure (a) illustrates the beam focused onto the top of the grating surface. Figures (b) and (c) show the beam focused on walls 1 and 2. The angle read on the rotation stage corresponds to the physical angle of walls 1 and 2 relative to "zero", i.e. the top grating surface.}
\end{figure} 

\subsection{Method 2}


For our second method, we position the grating on a rotation stage which is precise to 0.1 degrees, as illustrated in Fig \ref{fig:setup}. We then use the three discreet walls of the grating profile as reflectors and rotate the grating until the spot is reflected back in Littrow as illustrated in Fig \ref{fig:measurementwalls}. The rotation stage then reads the true angles subtended for a given reflecting wall.

To align the optical setup, we send Helium-Neon (HeNe) laser light to the part and look for the Littrow configuration by rotating the grating. Here we utilize two pinholes along the optical path to constrain the beam and align the two mirrors. After both mirrors are aligned, the pinholes are removed from the optical path and the initial beam and return beam should share the same path if the beam is incident-normal to the grating face (see Fig \ref{fig:setup}). When the spot returns along its incident path, it can be observed on a screen at the output of the laser. The grating is carefully attached to the precision rotation stage so that the axis of rotation is parallel to the etched rows. Between the rotation stage and grating is an x,y stage which assists in small changes in focus and positioning. 

With the beam incident perpendicular to the grating surface, we shift the grating with the x,y stage in the x-direction to focus the beam on the grating pattern (walls 1 or 2) and check for diffraction. With the x,y stage we then shift the grating such that the beam is focused on the top of the grating pattern (Fig \ref{fig:measurementwalls}a). With no diffraction from this flat surface, an un-diffracted beam should be seen returning directly along the initial beam path. If it is off-center, we make small adjustments on the stage to center the laser on the screen. Once centered, we have defined our zero on the rotation stage and can proceed to rotate the grating on the stage to take measurements of walls 1 and 2. Figures \ref{fig:measurementwalls}b and \ref{fig:measurementwalls}c show the orientation of the grating as viewed from above such that the beam is perpendicular to the surface of walls 1 and 2 respectively. By rotating to the angle which directs the beam directly back to the laser, the rotation stage reads the true value of the angle subtended for that wall. As in measurement method 1, these measurements are repeated multiple times and averaged to take the final value. The precision of this measurement, with 10 samples of the rotation stage angle, is taken to be $\frac{R}{2}= \frac{x_{max}-x_{min}}{2} \sim 0.2$ degrees.

In principle, a focused beam will have a symmetrical PSF after reflecting off a sufficiently large flat surface. In practice, it is challenging to focus the beam small enough to fit on just one wall of our gratings such that there is only reflection with no diffraction from neighboring grooves. Our ability to determine the center of the reflected spot (which we need to direct the return beam back along the exact initial beam path so that the rotation stage reads the real angle of the wall) is therefore limited by the size of the wall off which it is reflecting. Therefore, this method is more reliable for characterizing gratings which we have designed for longer wavelengths. Our gratings designed for the infrared L- and M-bands ($\sim$3 - 5.5 $\mu$m) have wider groove spacings (42.43$\mu$m and 64.9 $\mu$m respectively) and therefore wider groove faces than our gratings designed for the J-, H-, or K-bands ($\sim$1.1 - 2.5 $\mu$m). Wider groove faces provide a sufficiently large reflecting surface and therefore a more reliable angle determination. For L- and M-band gratings (i.e. best case),  the precision of the rotation stage limits the precision of this measurement method to $\sim 0.2$ degrees.

\subsection{Method 3}

\begin{figure}
\centering
\includegraphics[width=0.55\textwidth]{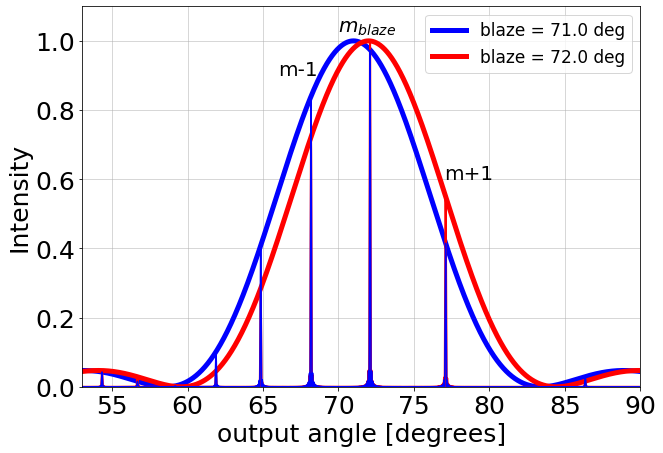}
\caption{\label{fig:offset} A blaze angle discrepancy of 1 degree would result in a small offset in the blaze envelope peak shown here. The series of peaks are the intensities of a monochromatic input signal diffracted into different orders and are limited in brightness by the blaze envelope. A blaze offset will shift the blaze envelope and have a measurable effect on the apparent intensity of blaze-adjacent orders. As shown above, for example, a blaze larger than expected will dim $m-1$ and will brighten $m+1$ (where $m=m_{blaze}$ is the brightest order). We can determine the fabricated blaze angle in comparison to the theoretical value from the discrepancies between in observed vs theoretical intensity of the $m+1$ and $m-1$ orders. Again, we note here that this model assumes incidence via free space, not immersion. Also, because of the geometry of our grating profile, groove width is smaller than groove spacing when not operated in immersion.}
\end{figure}

This method involves calculating the theoretical blaze function from the design blaze angle and comparing it to the true fabricated blaze function to discover the magnitude of any offsets. If our fabricated blaze is exactly what we expect, then the theoretical blaze envelope will exactly overlap with the blaze envelope obtained experimentally, and the intensities of each order will be the same in both cases. Otherwise, a non-zero discrepancy will result in an offset between the expected and experimentally obtained blaze envelopes, and discrepancies in brightness of the orders neighboring the blazed order. For an illustration of the offset and resulting varied order intensities, see Fig.\ref{fig:offset}. From observing the brightness of neighboring orders, we can determine the fabricated blaze angle in comparison to the theoretical value. We operate at very high order, so small deviations in the blaze angle will effect our blaze function drastically. We expect this method to yield a measurement precision of $\sim$0.1 degrees. It is important to note that this test is not carried out in immersion, meaning the groove width is not equal to the groove spacing for the propose of the test and blaze model. We assume light is incident on grating pattern via free space, in which case the groove width is only a fraction of the groove spacing. 

\begin{figure}[b] 
\centering
\includegraphics[width=1.05\textwidth]{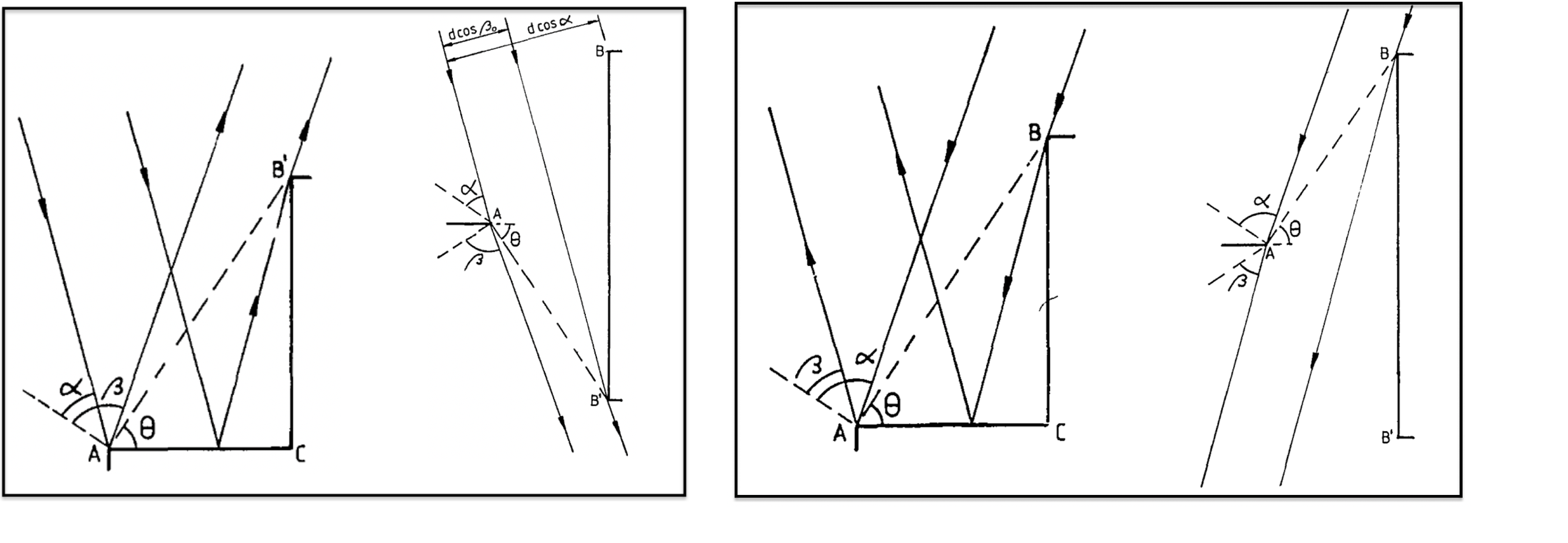}
\caption{\label{fig:modelslit} Above is a graphic taken from Engman 1982\cite{Engman:82} showing the relevant geometry for how a groove can be equivalently modeled as an offset slit where diffraction happens a the slit edges. The left diagrams show the groove geometry and equivalent slit model for the case $\alpha$\textless $\theta$. The right diagram shows the same for $\alpha$\textgreater $\theta$.  The `slit' is shown by the dashed line between groove edges A and B\textbf{'} for $\alpha$\textless $\theta$, and A and B for $\alpha$\textgreater $\theta$. There are two important consequences of this offset and tilted slit geometry. First, the incident plane wave only sees a projection of the slit. Second, there is a phase delay between the diffracted ray on one side of the beam relative to the other.}
\end{figure}

\subsubsection{Developing an Accurate Theoretical Model Description}

To accurately describe the expected blaze, we can model a groove as a slit that diffracts an incident plane wave. We understand optical systems to be Fourier transform machines; the transfer function that describes the input beam (i.e the slit/aperture) can be inverse Fourier transformed to get the impulse response. The impulse response squared describes the distribution of intensity on the output side of the system (Fig \ref{fig:modelslit} is an example of this). If the transfer function is a 2D slit described by a rect function (e.g. $E(x) = rect[\frac{x}{D}]$ for a slit size D), then the impulse response will be an airy disk described by a sinc function which depends on D and wavelength. This sinc function describes the intensity distribution in the airy disk and has the form $sinc = \frac{sin(x)}{x}$. A simple sinc function alone is descriptive of single slit diffraction, but modifications to the description must be made to account for the ways our ``slit'' is unconventional. Modifying the blaze envelope description with the appropriate geometrical descriptions has been done by many authors for various grating profiles and geometries (Engman\cite{Engman:87}, \cite{Engman:82}, Zhao\cite{doi:10.1080/09500349114552301} and others). We use these authors' models as a guide for developing our own blaze function definition.

A grating groove can be modeled as equivalent to a single slit, but in reflection instead of transmission. The incoming beam diffracts when it encounters groove edges, whereas no diffraction happens at the flat reflecting surfaces. This means we can ``unfold'' the system at the reflecting surface to create a slit with equivalent geometry (see Fig.\ref{fig:modelslit}). From this geometry, we see there are two distinct cases that we treat separately: $\alpha$\textgreater $\theta$ vs $\alpha$\textless $\theta$. In the case that $\alpha$\textgreater $\theta$, the diffraction happens before reflection, whereas if $\alpha$\textless $\theta$, the reflection happens after. 

\begin{figure} 
\centering
\includegraphics[width=0.98\textwidth]{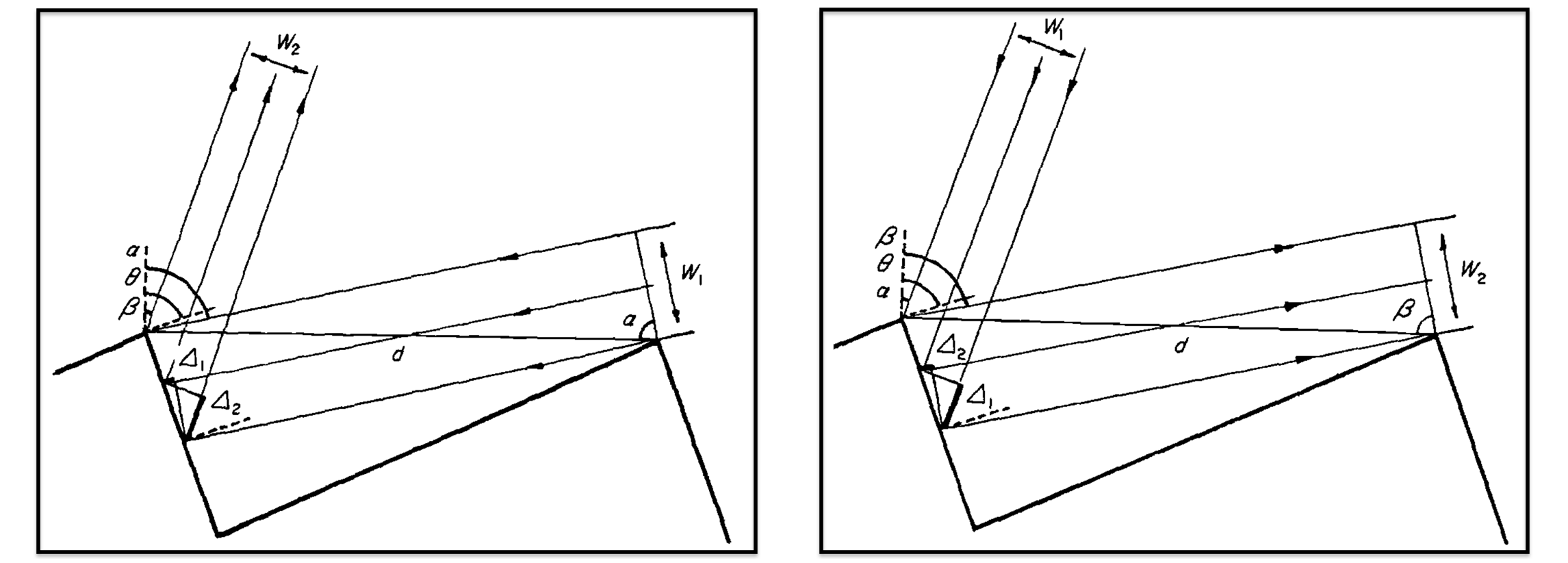}
\caption{\label{fig:Zhaomodel} Above is the geometry which defines a slightly different blaze model by F Zhao\cite{doi:10.1080/09500349114552301}. The left diagram is for $\alpha$\textless $\theta$ and the left is for $\alpha$\textgreater $\theta$. The phase lag is taken into account here via the beam width/projected slit size which is a function of angle of incidence.}
\end{figure}

One example of modified geometry can be found in Engman 1982\cite{Engman:82}, where the sinc function for $\alpha$\textless $\theta$ is defined as follows:

\begin{equation}
\label{eqn:Eng}
     I \sim \left[ \frac{sin\mathcal{H}}{\mathcal{H}} \left( \frac{cos\beta +cos(2\theta-\alpha)}{2cos\beta_{0}}\right) \right]^{2} 
\end{equation}

Where 
\begin{equation}
\label{eqn:ach}
    \mathcal{H} = \frac{\pi d}{\lambda} \left( sin(2\theta -\alpha) - sin\beta \right) 
\end{equation}

The sinc function $\frac{sin\mathcal{H}}{\mathcal{H}}$ has been modified to account for geometry shown in Fig \ref{fig:modelslit}. In eqn (\ref{eqn:Eng}), $cos\beta+cos(2\theta-\alpha)$ accounts for the skewedness of the slit relative to the incoming plane wave, and $2cos\beta_{0}$ is a normalizing factor 
which allows I = 1 at $\beta = \beta_{0} = 2\theta - \alpha$ for zeroth order (m = 0). Limitations of this particular model include the implied 90$^{\circ}$ groove valley angle, and groove shadowing is not necessarily accounted for. It is worth noting that Engman's model does include modifications for potential groove edge defect, which can be very useful for characterizing mechanically ruled gratings\cite{Engman:82}. Our chemically-etched gratings do not suffer from this effect because of the crystalline structure of silicon, but other gratings may benefit from analysis which includes this parameter.

Another version of a groove-slit model which is arguably closer to our desired description is given by Zhao\cite{doi:10.1080/09500349114552301} as:

\begin{equation}
     I = C^2 \left[\frac{sin\mathcal{U}}{\mathcal{U}}\right]^{2}
\end{equation}

where for $\alpha$\textless $\theta$:

\begin{equation}
    \mathcal{U} = \frac{\pi d}{\lambda}cos\beta \left(\frac{sin(\theta -\alpha)}{cos(\theta -\beta}+tan(\theta-\beta) \right)
\end{equation}

With the geometry shown in Fig \ref{fig:Zhaomodel}, this description takes into account the phase difference introduced by the skewedness of the slit and the projected slit width (i.e. the width of the beam).

Due to complications associated with the COVID-19 pandemic, our final theoretical envelope description is not yet complete or able to be verified in the lab. Next steps for this project involve iterating on a unique geometrical modification of the sinc function such that the output matches that of our grating in the lab.

\subsubsection{Experimental Process and Blaze Determination}

Once we have calculated the expected intensities via the blaze envelope function, we can calculate the expected output given our expected blaze, and compare to the experimentally determined intensities. To obtain this data experimentally, we use a similar optical set up as described in method 2, with the addition of a detector to allow us to measure intensities of the blaze adjacent orders. 

The method proposed by Engman\cite{Engman:82} involves measuring the relative intensities of orders $m_{\pm1}$ which neighbor the brightest order, and using those values to mathematically fit the true blaze (see Fig \ref{fig:offset} for an illustration of how a shifted blaze effects orders on either side of the brightest order). Comparing with the theoretical blaze model provides a direct measurement of how imprecise the blaze may be. We plan to utilize this method in the future. 

\section{Conclusion/Summary}

If the blaze of optical grating/grisms is off from the expected value, the overall instrument efficiency suffers. For transmission grisms, the wavelength of the undeviated array will also change slightly as a result of an imprecisely known blaze. Precisely measuring the blaze of immersion gratings and grisms is critical to preserving instrument efficiency. We have detailed multiple methods to measure the fabricated blaze of our in-house silicon gratings. By precisely measuring the blaze using these methods, we can quantify the accuracy with which wet etching produces the required blaze, and further optimize the process for overall grating efficiency. 

\appendix    

\acknowledgments 

We would like to thank Sierra Hickman for her previous work on these measurement methods on which we continue to build. We would also like to thank Benjamin Kidder for his guidance and previous work on the fabrication and characterization of our immersion gratings and characterization techniques. 

This work was supported through NASA APRA grant NNX15AD97G, and by the Giant Magellan Telescope Organization and Jet Propulsion Laboratory. This research was supported by the Jet Propulsion Laboratory, California Institute of Technology, under a contract with the National Aeronautics and Space Administration and funded through the internal Research and Technology Development program.

\bibliography{report} 

\begin{thebibliography}{10}

\bibitem{10.1117/12.317263}
Jaffe, D.~T., Keller, L.~D., and Ershov, O.~A., ``{Micromachined silicon
  diffraction gratings for infrared spectroscopy},'' in [{\em Infrared
  Astronomical Instrumentation}{\nolinebreak\hspace{0.1em}]},  Fowler, A.~M.,
  ed.,  {\bf 3354},  201 -- 212, International Society for Optics and
  Photonics, SPIE (1998).

\bibitem{10.1117/12.857164}
Wang, W., Gully-Santiago, M., Deen, C., Mar, D.~J., and Jaffe, D.~T.,
  ``{Manufacturing of silicon immersion gratings for infrared spectrometers},''
  in [{\em Modern Technologies in Space- and Ground-based Telescopes and
  Instrumentation}{\nolinebreak\hspace{0.1em}]},  Atad-Ettedgui, E. and Lemke,
  D., eds.,  {\bf 7739},  1575 -- 1583, International Society for Optics and
  Photonics, SPIE (2010).

\bibitem{10.1117/12.2057023}
Ge, J., Powell, S., Zhao, B., Schofield, S., Varosi, F., Warner, C., Liu, J.,
  Sithajan, S., Avner, L., Jakeman, H., Gittelmacher, J.~A., Yoder, W.~A.,
  Muterspaugh, M., Williamson, M., and Maxwell, J.~E., ``{On-sky performance of
  a high resolution silicon immersion grating spectrometer},'' in [{\em
  Ground-based and Airborne Instrumentation for Astronomy
  V}{\nolinebreak\hspace{0.1em}]},  Ramsay, S.~K., McLean, I.~S., and Takami,
  H., eds.,  {\bf 9147},  486 -- 497, International Society for Optics and
  Photonics, SPIE (2014).

\bibitem{10.1117/12.2232064}
Rayner, J., Tokunaga, A., Jaffe, D., Bonnet, M., Ching, G., Connelley, M.,
  Kokubun, D., Lockhart, C., and Warmbier, E., ``{iSHELL: a construction,
  assembly and testing},'' in [{\em Ground-based and Airborne Instrumentation
  for Astronomy VI}{\nolinebreak\hspace{0.1em}]},  Evans, C.~J., Simard, L.,
  and Takami, H., eds.,  {\bf 9908},  2401 -- 2417, International Society for
  Optics and Photonics, SPIE (2016).

\bibitem{10.1117/12.2232994}
Jaffe, D.~T., Barnes, S., Brooks, C., Lee, H., Mace, G., Pak, S., Park, B.-G.,
  and Park, C., ``{GMTNIRS: progress toward the Giant Magellan Telescope
  near-infrared spectrograph},'' in [{\em Ground-based and Airborne
  Instrumentation for Astronomy VI}{\nolinebreak\hspace{0.1em}]},  Evans,
  C.~J., Simard, L., and Takami, H., eds.,  {\bf 9908},  648 -- 656,
  International Society for Optics and Photonics, SPIE (2016).

\bibitem{10.1117/12.857127}
Keller, L., Deen, C.~P., Jaffe, D.~T., Ennico, K.~A., Greene, T.~P., Adams,
  J.~D., Herter, T., and Sloan, G.~C., ``{Progress report on FORCAST grism
  spectroscopy as a future general observer instrument mode on SOFIA},'' in
  [{\em Ground-based and Airborne Instrumentation for Astronomy
  III}{\nolinebreak\hspace{0.1em}]},  McLean, I.~S., Ramsay, S.~K., and Takami,
  H., eds.,  {\bf 7735},  2406 -- 2411, International Society for Optics and
  Photonics, SPIE (2010).

\bibitem{10.1117/12.857568}
Gully-Santiago, M., Wang, W., Deen, C., Kelly, D., Greene, T.~P., Bacon, J.,
  and Jaffe, D.~T., ``{High-performance silicon grisms for 1.2-8.0 um: detailed
  results from the JWST-NIRCam devices},'' in [{\em Modern Technologies in
  Space- and Ground-based Telescopes and
  Instrumentation}{\nolinebreak\hspace{0.1em}]},  Atad-Ettedgui, E. and Lemke,
  D., eds.,  {\bf 7739},  1353 -- 1359, International Society for Optics and
  Photonics, SPIE (2010).

\bibitem{10.1117/12.869018}
van Amerongen, A.~H., Visser, H., Vink, R. J.~P., Coppens, T., and Hoogeveen,
  R. W.~M., ``{Development of immersed diffraction grating for the TROPOMI-SWIR
  spectrometer},'' in [{\em Sensors, Systems, and Next-Generation Satellites
  XIV}{\nolinebreak\hspace{0.1em}]},  Meynart, R., Neeck, S.~P., and Shimoda,
  H., eds.,  {\bf 7826},  345 -- 352, International Society for Optics and
  Photonics, SPIE (2010).

\bibitem{10.1117/12.2309092}
van Amerongen, A., Krol, H., Grèzes-Besset, C., Coppens, T., Bhatti, I., Lobb,
  D., Hardenbol, B., and Hoogeveen, R., ``{State of the art in silicon immersed
  gratings for space},'' in [{\em International Conference on Space Optics —
  ICSO 2012}{\nolinebreak\hspace{0.1em}]},  Cugny, B., Armandillo, E., and
  Karafolas, N., eds.,  {\bf 10564},  713 -- 719, International Society for
  Optics and Photonics, SPIE (2017).

\bibitem{Sokal_2020}
Sokal, K.~R., Johns-Krull, C.~M., Mace, G.~N., Nofi, L., Prato, L., Lee, J.-J.,
  and Jaffe, D.~T., ``The mean magnetic field strength of {CI} tau,'' {\em The
  Astrophysical Journal}~{\bf 888},  116 (jan 2020).

\bibitem{Flagg_2019}
Flagg, L., Johns-Krull, C.~M., Nofi, L., Llama, J., Prato, L., Sullivan, K.,
  Jaffe, D.~T., and Mace, G., ``{CO} detected in {CI} tau b: Hot start implied
  by planet mass and m k,'' {\em The Astrophysical Journal}~{\bf 878},  L37
  (jun 2019).

\bibitem{Johns_Krull_2016}
Johns-Krull, C.~M., McLane, J.~N., Prato, L., Crockett, C.~J., Jaffe, D.~T.,
  Hartigan, P.~M., Beichman, C.~A., Mahmud, N.~I., Chen, W., Skiff, B.~A.,
  Cauley, P.~W., Jones, J.~A., and Mace, G.~N., ``A {Candidate} {Young}
  {Massive} {Planet} {in} {Orbit} {Around} {The} {Classical} {T} {Tauri} {Star}
  {CI} {Tau},'' {\em The Astrophysical Journal}~{\bf 826},  206 (aug 2016).

\bibitem{10.1117/12.672198}
Marsh, J.~P., Mar, D.~J., and Jaffe, D.~T., ``{Fabrication and performance of
  silicon immersion gratings for infrared spectroscopy},'' in [{\em
  Ground-based and Airborne Instrumentation for
  Astronomy}{\nolinebreak\hspace{0.1em}]},  McLean, I.~S. and Iye, M., eds.,
  {\bf 6269},  1459 -- 1470, International Society for Optics and Photonics,
  SPIE (2006).

\bibitem{10.1117/12.2057468}
Brooks, C.~B., Gully-Santiago, M., Grigas, M., and Jaffe, D.~T., ``{New
  metrology techniques improve the production of silicon diffractive optics},''
  in [{\em Advances in Optical and Mechanical Technologies for Telescopes and
  Instrumentation}{\nolinebreak\hspace{0.1em}]},  Navarro, R., Cunningham,
  C.~R., and Barto, A.~A., eds.,  {\bf 9151},  461 -- 470, International
  Society for Optics and Photonics, SPIE (2014).

\bibitem{10.1117/12.2314271}
Kidder, B.~T., Brooks, C.~B., Grigas, M.~M., Hickman, S., and Jaffe, D.~T.,
  ``{Manufacturing silicon immersion gratings on 150-mm material},'' in [{\em
  Advances in Optical and Mechanical Technologies for Telescopes and
  Instrumentation III}{\nolinebreak\hspace{0.1em}]},  Navarro, R. and Geyl, R.,
  eds.,  {\bf 10706},  589 -- 595, International Society for Optics and
  Photonics, SPIE (2018).

\bibitem{IMGJ}
Abramoff, M., Magalhaes, P., and Ram, S., ``Image processing with imagej,''
  {\em Biophotonics International}~{\bf 11}(7),  36--42 (2004).

\bibitem{Engman:82}
Engman, S. and Lindblom, P., ``Blaze characteristics of echelle gratings,''
  {\em Appl. Opt.}~{\bf 21},  4356--4362 (Dec 1982).

\bibitem{Engman:87}
Engman, S., Lindblom, P., and Olsson, B., ``Testing echelle gratings: a simple
  method,'' {\em Appl. Opt.}~{\bf 26},  26--28 (Jan 1987).

\bibitem{doi:10.1080/09500349114552301}
F.Zhao, ``A diffraction model for echelle gratings,'' {\em Journal of Modern
  Optics}~{\bf 38}(11),  2241--2246 (1991).

\end{thebibliography}
\bibliographystyle{spiebib} 

\end{document}